\documentclass[useAMS,usenatbib]{mn2e} 
\usepackage{aas_macros}
\usepackage{graphics}
\usepackage{epsfig}  
\usepackage{natbib} 
\usepackage{color}
\usepackage{float}
\usepackage{amsmath}
\usepackage{times}
\usepackage{upgreek}
\usepackage[varg]{txfonts}
\newcommand{\Tab}[1]{Table~\ref{#1}}
\newcommand{\Sec}[1]{Section~\ref{#1}}
\newcommand{\Eq}[1]{Eq.(\ref{#1})}
\newcommand{\Fig}[1]{Fig.~\ref{#1}}

\newcommand{\hMpc}{{\ifmmode{h^{-1}{\rm Mpc}}\else{$h^{-1}$Mpc}\fi}}
\newcommand{\hkpc}{{\ifmmode{h^{-1}{\rm kpc}}\else{$h^{-1}$kpc}\fi}}
\newcommand{\hMsun}{{\ifmmode{h^{-1}{\rm {M_{\odot}}}}\else{$h^{-1}{\rm{M_{\odot}}}$}\fi}}
\newcommand{\ltsima}{$\; \buildrel < \over \sim \;$}
\newcommand{\gtsima}{$\; \buildrel > \over \sim \;$}
\newcommand{\lsim}{\lower.5ex\hbox{\ltsima}}
\newcommand{\gsim}{\lower.5ex\hbox{\gtsima}}

\def\lesssim{\mathrel{\hbox{\rlap{\hbox{\lower4pt\hbox{$\sim$}}}\hbox{$<$}}}}
\def\gtrsim{\mathrel{\hbox{\rlap{\hbox{\lower4pt\hbox{$\sim$}}}\hbox{$>$}}}}

\title[The impact of baryonic physics on subhaloes]
      {The impact of baryonic physics on the shape and radial alignment of substructures in cosmological dark matter haloes}
\author[Knebe et al.] 
{Alexander Knebe$^1$, Noam I Libeskind$^2$, Steffen R. Knollmann$^1$, Gustavo Yepes$^1$, \newauthor Stefan Gottl\"ober$^2$, Yehuda Hoffman$^3$
  \\
  $^1$Grupo de Astrof\'\i sica, Departamento de Fisica Teorica, Modulo C-15, Universidad Aut\'onoma de Madrid, Cantoblanco E-28049, Spain\\
  $^2$Astrophysikalisches Institut Potsdam, An der Sternwarte 16, D-14482 Potsdam, Germany\\
 $^3$Racah Institute of Physics, The Hebrew University of Jerusalem, Givat Ram, Israel
  }

\begin{document}

\date{First draft}

\pagerange{\pageref{firstpage}--\pageref{lastpage}} \pubyear{2008}

\maketitle

\label{firstpage}

\begin{abstract}
We use two simulations performed within the Constrained Local UniversE Simulation (CLUES) project to study both the shape and radial alignment of (the dark matter component of) subhaloes; one of the simulations is a dark matter only model while the other run includes all the relevant gas physics and star formation recipes. We find that the involvement of gas physics does not have a statistically significant effect on either property -- at least not for the most massive subhaloes considered in this study. However, we observe in both simulations including and excluding gasdynamics a (pronounced) evolution of the dark matter shapes of subhaloes as well as of the radial alignment signal since infall time. Further, this evolution is different when positioned in the central and outer regions of the host halo today; while subhaloes tend to become more aspherical in the central 50\% of their host's virial radius, the radial alignment weakens in the central regime while strengthening in the outer parts. We confirm that this is due to tidal torquing and the fact that subhaloes at pericentre move too fast for the alignment signal to respond.

\end{abstract}


\section{Introduction}
\label{sec:introduction}
Over the past decade there has been a tremendous increase in activity dedicated to satellite galaxies - the small, over dense, clumps of matter that inhabit dark matter haloes. Specifically, the study of satellite properties has been attacked from both a numerical perspective  \citep[e.g.][]{Gao04,Kravtsov04a, Gill04b,Gill05,Diemand07b,Kuhlen07,Pereira08,Warnick06,Sales07,Libeskind07,Warnick08,Springel08,Klimentowski10,Elahi09,Ludlow09,Tissera09,Dolag09} as well as an observational one \citep[e.g.][]{Azzaro06,Faltenbacher07,Koposov08,Wang08,Bailin08,Chakrabarti09,Wang09}. The advent of ``Near-Field Cosmology'' \citep[cf.][]{Freeman02} has opened a new field dedicated to scrutinizing the (commonly accepted) standard model of comology that is based on the existence of dark matter and dark energy \citep[e.g.][]{Komatsu09}. Characteristics that were previously beyond modeling include, for example, the shapes, orientation, and alignment of subhaloes \citep{Knebe04,Zentner05,Libeskind05,Kuhlen07,Knebe08,Faltenbacher08,Pereira08}. Furthermore, while there is a clear consensus that baryonic physics will have an effect on structure formation, especially on galactic and sub-galactic scales that are only now within reach of simulations \citep{Maccio06, Weinberg08, Okamoto09,Tissera09,Dolag09, Libeskind10}, a rigorous quantification of these effects has not yet been fully completed.

In this paper we use a set of two self-consistent cosmological simulations of the formation of the Local Group of galaxies\footnote{See the ``Constrained Local UniversE Simulations' (CLUES) project \texttt{http://www.clues-project.org} for more details.}. In one simulation we  focus just on the gravitational interaction of dark matter whereas the second run incorporates baryonic physics as well. The simulations are  studied with respects to their (dark matter) shapes as well as the so-called radial alignment of subhaloes: this is the tendency of subhaloes to have their intrinsic shapes aligned towards the centre of their respective host halo. While such a phenomenon has been reported on a broad range of scales including massive clusters as well as galactic systems \citep{Hawley75, Thompson76,Pereira05,Agustsson06,Wang07,Faltenbacher08} the literature still lacks the confirmation of such a signal in gasdynamical simulations of cosmic structure formation. Further, it has been argued that the shapes of (field) dark matter haloes can be modified by galaxy formation \citep[e.g.][]{Katz91b,Kazantzidis04,Bailin05b,Maccio06,Weinberg08,Libeskind10} even though the consequence  may be quite small for dwarf galaxies \citep[cf.][]{Diemand09} and subhaloes \citep[cf.][]{Kazantzidis04b}. It therefore remains interesting to check whether or not there is a measurable effect for subhaloes and if the predicted (and observed) radial alignment signal persist when including such physics in the simulations.

Please note that in this paper we investigate the dark matter shapes and the influence of baryonic physics on them; an in-depth study of the shapes of the baryonic component will be presented in a companion paper (Libeskind et al., in preparation).

\section{The Simulations}
\label{sec:simulations}
In this Section we describe the simulations used throughout this study and the methodology employed to identify host haloes and their substructure.

\subsection{Constrained Simulations of the Local Group}
\label{sec:localgroup}

\begin{table}
\begin{center}
 \begin{tabular}{l l l l l l}
      &$m_{\rm gas,i}$  &  $m_{\rm dm}$  & $\epsilon$ & $N_{\rm dm}$ \\
   \hline
   \hline
   DM: & $ - $ & $2.54  \times 10^{5}$ & $0.15$ & $5.29\times10^{7}$  \\
   
   SPH : & $4.42 \times 10^{4}$ & $2.1 \times 10^{5}$ & $0.15$ &
   $5.29\times10^{7}$  \\

    \hline

 \end{tabular}
 \end{center}
\caption{ Simulation parameters. From left to right, the initial mass per gas
  particle (in units of $h^{-1} \rm M_{\odot}$); mass of high resolution dark
  matter particles (in units of $h^{-1} \rm M_{\odot}$); softening length (in units of $h^{-1} \rm kpc$); Number of
  high resolution DM particles.}
\label{table1}
\end{table}

In this paper, we use the same set of simulations as those published in \citet{Libeskind10} and refer the reader to that paper for a more exhaustive description and discussion of the various aspects of the simulations including the constraints that were imposed in order to reproduce a $z=0$ Local Group. As mentioned earlier, these simulations form part of  the CLUES project. We briefly summarize the main properties of the simulation here, for clarity.

We choose to run our simulations using standard $\Lambda$CDM initial conditions, that assume a WMAP3 cosmology \citep{Spergel07}, i.e.  $\Omega_m = 0.24$, $\Omega_{b} = 0.042$, $\Omega_{\Lambda} = 0.76$. We use a normalization of $\sigma_8 = 0.73$ and a $n=0.95$ slope of the power spectrum. We used the PMTree-SPH MPI code \texttt{GADGET2} \citep{Springel05} to simulate the evolution of a cosmological box with side length of $L_{\rm box}=64 h^{-1} \rm Mpc$. Within this box we identified (in a lower-resolution run) the position of a model local group that closely resembles the real Local Group \citep[cf.][]{Libeskind10} and re-simulated a $2 h^{-1} \rm Mpc$ region about the centre of this local group using our full resolution equivalent to $4096^3$ effective particles. Structures that are still linear at $z=0$ are constrained to have various properties that are similar to objects in the local Universe, allowing us to simulate our host halos in an environment similar to the real one.

We use the same set of initial conditions to run two simulations, one with dark matter only and another one with dark matter, gas dynamics, cooling, star formation and supernovae feedback. We shall call these two simualtions DM and SPH, respectively; the numerical parameters of our simulations are summarized in \Tab{table1}.

For the gas dynamical SPH simulation, we follow the feedback and star formation rules of \cite{Springel03}: the interstellar medium (ISM) is modeled as a two phase medium composed of hot ambient gas and cold gas clouds in pressure equilibrium. The thermodynamic properties of the gas are computed in the presence of a uniform but evolving ultra-violet cosmic background generated from QSOs and AGNs and switched on at $z=6$ \citep{Haardt96}.  Cooling rates are calculated from a mixture of a primordial plasma composition. No metal dependent cooling is assumed, although the gas is metal enriched due to supernovae explosions. Molecular cooling below $10^{4} {\rm K}$ is also ignored.  Cold gas cloud formation by thermal instability, star formation, the evaporation of gas clouds, and the heating of ambient gas by supernova driven winds are assumed to all occur simultaneously.

\subsection{The (Sub-)Halo Finding}
\label{sec:halofinding}
In order to identify halos and subhalos in our simulation we have run the MPI+OpenMP hybrid halo finder \texttt{AHF}\footnote{\texttt{AMIGA} halo finder, to be downloaded freely from \texttt{http://www.popia.ft.uam.es/AMIGA}} described in detail in \cite{Knollmann09}. \texttt{AHF} is an improvement of the \texttt{MHF} halo finder \citep{Gill04a}, which locates local overdensities in an adaptively smoothed density field as prospective halo centres. The local potential minima are computed for each of these density peaks and the gravitationally bound particles are determined. Only peaks with at least 20 bound particles are considered as haloes and retained for further analysis. We would like to stress that our halo finding algorithm automatically identifies haloes, sub-haloes, sub-subhaloes, etc. For more details on the mode of operation and actual functionality we refer the reader to the code description paper by \citet{Knollmann09}.

For each halo, we compute the virial radius $R_{\rm halo}$, that is the radius $r$ at which the density $M(<r)/(4\pi r^3/3)$ drops below $\Delta_{\rm  vir}\rho_{\rm b}$. Here $\rho_{\rm b}$ is the cosmological background matter density. The threshold $\Delta_{\rm vir}$ is computed using the spherical top-hat collapse model and is a function of both cosmological model and time. For the cosmology used in this paper, $\Delta_{\rm vir}=355$ at $z=0$. Subhaloes are defined as haloes which (fully) lie within the virial region of a more massive halo, the so-called host halo. As subhaloes are embedded within the density of their respective host halo, their own density profile usually shows a characteristic upturn at a radius $R_t \lsim R_{\rm halo}$, where $R_{\rm halo}$ would be their actual (virial) radius if they were found in isolation.\footnote{Please note that the actual density profile of subhaloes  after the removal of the host's background drops faster than for isolated  haloes \citep[e.g.][]{Kazantzidis04}; only when measured against the main halo's background will we find the characteristic upturn used here to  define the truncation radius $r_t$.}  We use this ``truncation radius'' $R_t$ as the outer edge of the subhalo (and refer to it as  $R_{\rm sat}$) and hence subhalo properties (i.e. mass, density profile, velocity dispersion, rotation curve) are calculated using the gravitationally bound particles inside the truncation radius $R_t$. For a host halo we calculate properties using the virial radius $R_{\rm halo}$ referred to as $R_{\rm host}$.

We construct merger trees by cross-correlating haloes in consecutive simulation outputs. For this purpose, we use a tool that comes with the \texttt{AHF} package, called \texttt{MergerTree} that follows both host and subhalos identified at redshift $z=0$ backwards in time. At a previous snapshot, the code identifies as the direct progenitor the halo that shares the most particles with the present halo \textit{and} is closest in mass. Again, for more elaborate details we point the reader to \citet{Libeskind10}.

These merger trees are in turn used to define the infall time of a subhalo, i.e. in our case the time it is fully embedded within the virial radius of its host: $d_{\rm sat}$+$R_{\rm sat}<R_{\rm host}$. The accuraacy of the determination of this infall time is sensitive to the time sampling of our outputs which are equally spaced in $\Delta t\approx100$Myrs for the DM model and $\Delta t\approx30$Myrs for the SPH model, respectively.

\section{Results}
\label{sec:results}
\begin{table}
\begin{center}
 \begin{tabular}{l c c c c}
      & \multicolumn{1}{l}{shape study:} & \multicolumn{1}{l}{radial alignment study:}\\
      & $M_{\rm sub} > 2\times 10^{8}h^{-1} M_{\odot}$  &   $M_{\rm sub} > 2\times 10^{8}h^{-1} M_{\odot}$ \& $b/a<0.9$ \\
  \hline
   \hline
   DM: &  64 = 17 + 47 & 56 = 16 + 40\\
   SPH: & 45 = 20 + 25 &  36 = 18 + 18\\
    \hline
 \end{tabular}
 \end{center}
\caption{The number of subhaloes in each model according to the various selection criteria. The splitting into two summands gives the number of subhaloes closer (first number) and farther away (second number) than 50\% of the host's virial radius.}
\label{table2}
\end{table}
 
Using our two simulations and their respective (sub-)halo analysis we study the shapes and the orientation of the shapes of subhaloes with respect to their position within the host halo. We restrict this analysis to the three most massive host haloes, namely the Milky Way, M31, and M33, and stack all subhaloes found in these hosts in all analysis and plots.  In order to compare like with like, we make a mass cut, comparing only the most massive of the $z=0$ subhaloes: following \citet{Libeskind10} -- who studied the same set of simulations with respect to the radial distribution of subhaloes -- we consider only subhalos with masses larger than $M_{\rm sub} > 2\times 10^{8}h^{-1} M_{\odot}$, which roughly corresponds to subhaloes with more than 1000 particles. This cut was chosen in order to obtain the correct number of visible satellites orbiting about the Milky Way ($\sim 20$). Further, for the analysis of the radial alignment in Sec.~\ref{sec:radialalignment} we wish to study satellites with well defined shapes and thus employ the additional criterion $b/a<0.9$ (with $a>b>c$ being the eigenvalues of the moment of inertia tensor) as suggested (and successfully applied) in \citet[e.g.][]{Pereira08} and \citet{Knebe08}. The respective numbers of subhaloes left for the analysis are summarized in \Tab{table2}. Note that we also split the sample according to whether the subhalo is within half the host's virial radius ($d_{\rm sat} < 0.5~R_{\rm host}$) or not ($0.5~R_{\rm host}<d_{\rm sat}<R_{\rm host}$); the number of subhalos falling into these two bins is also shown in \Tab{table2}.

\subsection{The Shapes of Subhaloes}
\label{sec:shapes}

\subsubsection{Determining the shape} \label{sec:howto}
\begin{figure}
\noindent
\centerline{\hbox{\psfig{figure=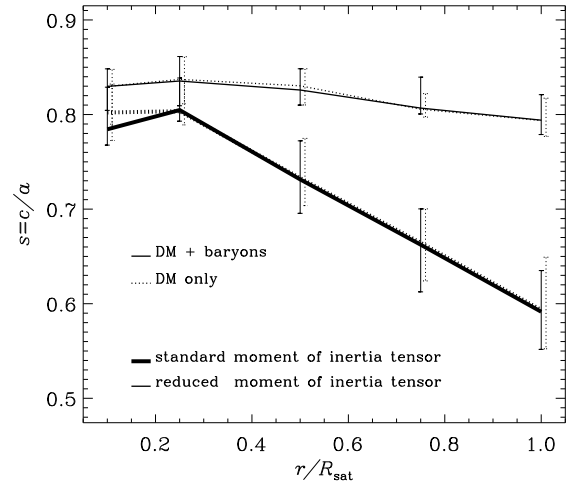,width=\hsize,angle=0}}}
\caption{The radial distribution of the medians of the sphericity $s=c/a$ for all particles (solid black) and the dark matter particles only (dotted black) in the SPH model. The thick lines correspond to $a>b>c$ derived from the standard moment of inertia tensor while the thin lines utilized the reduced moment of inertia tensor. The error bars are the upper and lower quartile of the plotted medians.}
\label{fig:shapeBC}
\end{figure}

Before quantifying the shape and the radial alignment of shapes, we first must confirm that the subhaloes we identify, are in fact aspherical. Several options for calculating a subhalo's shape exist which we wish to investigate. To this extent we calculate the subhaloes' sphericity $s=c/a$, where $a>b>c$ are the eigenvalues of the moment of inertia tensor of the subhalo.

There are two commonly used definitions for this tensor. These are the (standard) moment of inertia tensor

\begin{equation}
I_{i,j}=\sum_nm_n x_{i,n}x_{j,n}
\end{equation}

\noindent
and the reduced moment of inertia tensor

\begin{equation}
I_{i,j}^{\rm red}=\sum_nm_n x_{i,n}x_{j,n}/r_n^2 .
\end{equation}

\noindent
where in both cases $x_{i,n}$ and $x_{j,n}$ are the $i$th and $j$th component of the $n$th particle coordinate and $r_n$ its distance to the subhalo centre; note that the reduced version puts more weight onto the central region; or ino ther words, the standard moment of intertia tensor weights each particle's contribution by $r_{\rm n}^2$ which is compensated for by the $1/r_{\rm n}^2$ factor in the reduced one. We calcluate both $I_{i,j}$ and $I_{i,j}^{\rm red}$ using all (bound) particles within the radius $R_{\rm sat}$ of the subhalo (or interior to some radius $r<R_{\rm sat}$ when studying the radial shape profile). Therefore, our determination of the eigenvalues used for the shape determination is based upon a diagonolisation of the moment of intertia tensor based upon particles within a spherical sphere. This approach biases the shapes towards higher sphericities. As shown by \citet{Bailin05} this can be corrected for, i.e. $(c/a)=(c/a)_{\rm spherical}^{\sqrt{3}}$. We apply this correction to all shapes and refer to these corrected shapes as $c/a$ throughout the paper. Note that using spherical shells rather than the ellipsoids does not affect the orientation of the principal axes and is hence irrelevant for the study of the radial alignment.

In order to properly calculate subhalo shape we wish to determine how these two definitions for the moment of Inertia Tensors  differ - not only between each other but also, for the SPH run, when considering all subhalo particles or just the dark matter component. In \Fig{fig:shapeBC}  we show the median sphericity $s=c/a$, calculated using the two moment of inertia tensors, for the dark matter subhalos as well as for all particles in the SPH subhalos and the dark matter component of SPH subhalos\footnote{Note that we are not utilizing the shape of the baryonic component in this particular study; we are mainly interested in the (possible) differences in the dark matter component arising from the inclusion of gas physics into the cosmological simulation.}. The difference in the two definitions is clearly seen: the reduced moment of inertia tensor (thin lines) is dominated by the inner region distribution well towards the edge of the subhalo; that is, the sphericity remains nearly constant at the peak value reached at approximately $0.1R_{\rm sat}$. We further observe that for the SPH subhalos, the distributions for the total (solid back) and dark matter (dashed black) are practically indistinguishable, leading to the conclusion that baryons play a minor role in the shape determination of the subhalo -- at least in the outer parts $r/R_{sat}\geq0.1$. Despite the quality of our simulations we are still unable to resolve the very inner parts of these subhaloes. However, we also observe (though not shown here) the drop towards stronger asphericity in the very central regions as previously found by \citet{Kuhlen07} in their dark matter only simulations. We additionally like to mention (also not explicitly shown here) that the drop in sphericity towards the subhalo edges (seen when utilizing the standard moment of inertia tensor) is driven by subhaloes closer to the host's centre as  these particular subhaloes feel stronger tidal forces. 

In \Fig{fig:shapeBC} we showed the difference in the shape of subhaloes using either all particles or just the dark matter particles \textit{in the SPH run alone}. We are therefore unable to draw any conclusions regarding the affect of baryons on an individual (subhalo's) shape, since to do this would require comparing the difference in shape between a given DM  subhalo and the same subhalo modeled with  SPH. However \Fig{fig:shapeBC} shows that it makes little (if any) difference if shapes are calculated  using all or just the dark matter particles in the SPH model. 

However, one may rightfully raise the question whether or not our results are mere numerical artifacts or refer to a physical phenomenon. To shed some light on this issue we checked for two things: first, do we actually resolve the (inner parts of the) subhaloes with sufficient baryonic particles; and second, are our results biased by the halo finder and the position of the subhalo within the host, respectively.

In order to understand the lack of difference in the shapes when using either the total or dark matter only component in the SPH model, we need to examine the baryon fraction. We can confirm that the median ratio of number of baryonic particles to the total number of particles within $\approx 0.1 R_{\rm sat}$ is close to 35\%. This ratio drops to about 20\% at the subhalo's outer edge indicating that there is in fact a sufficient number of baryonic particles present throughout the subhaloes' volume. Note that this test confirms the presence of baryonic particles; their contribution to the shape/sphericitiy is further dependent on the mass (not number) fraction due to the weights $m_i$ in the definition both moment of inertia tensors, i.e. the actual baryonic mass fractions at the aforementioned radii are $f_b=0.09$ at $0.1R_{\rm sat}$ and $f_b=0.055$ at the outer subhalo edge, respectively.

Additionally, our spherical overdensity halo finder \texttt{AHF} biases the shape determination owing to the fact that subhaloes closer to the host centre will have smaller radii \citep[cf. also discussion in Section 3.2 in][]{Knebe08}. This is due to the embedding of subhaloes within a sea of host particles such that objects closer to the central density peak of the host will be truncated earlier leading to a smaller radiues. To check whether or not our sample is affected by this effect we studied the relation between subhalo radius $R_{\rm sat}$ and position $d_{\rm sat}$ within the host. We do not find any significant correlation for the subhalo sample under investigation with the Spearman rank coefficient of the two quantities being 0.11. To gauge the value of the Spearman rank coefficient we tested several pure random correlations consisting of the same number of particles as subhaloes studied here and found the mean Spearman rank coefficient for these random samples to be 0.10; hence our value of 0.11 suggests no correlation.\footnote{The Spearman rank coefficient is a non-parametric measure of correlation: it assesses how well an arbitrary monotonic function describes the relationship between two variables, without making any other assumptions about the particular nature of the relationship between the variables \citep{Kendall90}.}

Lingering on the results obtained so far for a moment leads to the following conclusions: since the standard moment of inertia tensor is unbiased by the central region and thus more sensitive to (radial) shape variations, we will use this measure rather than the reduced moment of inertia tensor throughout this work. Or put differently, we are interested in the shape variation as a function of radius $r/R_{\rm sat}$ where the shape at a particular radius should be calculted utilizing a shell of thickness $[r/R_{\rm sat}-\epsilon,r/R_{\rm sat}+\epsilon]$. However, despite the quality of our simulations there are still not enough particles within such shells for the subhaloes and hence we prefer to employ the standard moment of inertia tensor that actually puts more weight onto the particles in the region close to $r/R_{\rm sat}$. Further, as baryons have no effect on the subhalo shape determination, we use all subhalo particles for the shape determination in both our models (i.e. DM and SPH).

We wish to remind the reader that up to now none of the results suggests that the baryons do not have an influence on the shape as this requires a direct comparison of the DM vs. SPH model (see Subsection~\ref{sec:shaperad}).

\subsubsection{Radial profile of host sphericity} \label{sec:hostshapes}
\begin{figure}
\noindent
\centerline{\hbox{\psfig{figure=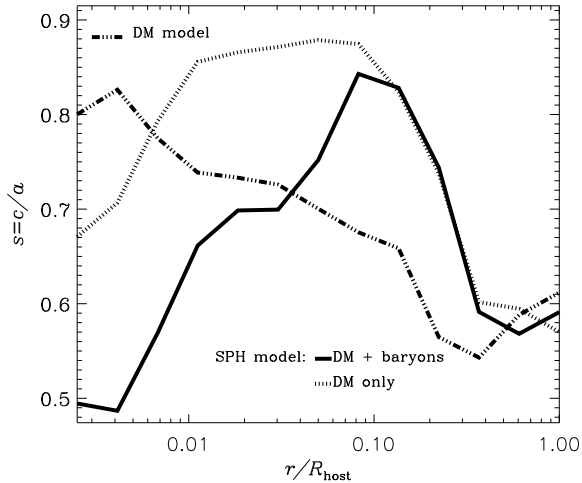,width=\hsize,angle=0}}}
\caption{The mean radial distribution of the sphericity of our three host haloes in the DM (dot-dashed line) and SPH model (solid and dashed lines). The measurements are based upon the standard moment of inertia tensor and for the SPH model upon the total (solid line) and DM only component (dashed line).}
\label{fig:shapesChosts}
\end{figure}

Even though we primarily address subhalo shape, we nevertheless like to also present the (mean) radial sphericity profile for the three host haloes under investigation, too; this allows us to get a better feeling and gauge for the differences between subhaloes and host haloes, respectively. This subject has been the target of many previous investigations by other groups \citep[e.g.][]{Dubinski94,Gustafsson06, Debattista08, Abadi09, Tissera09} with the conclusion that the inclusion of baryons drive the dark matter halo to become more spherical.

Our results (based upon the standard moment of inertia tensor) can be viewed in \Fig{fig:shapesChosts} where we show the sphericity for the DM model (dot-dashed line) alongside the two profiles, i.e. DM only (dashed line) and total matter (solid line), for the SPH model. We clearly see differences in the sphericities in the SPH model below $0.1 R_{\rm host}$ in the sense that the dark matter component alone appears more spherical than the total matter. This transition at $0.1 R_{\rm host}$ coincides with the value reported by \citet{Bailin05b} who found that below this scale the inner halo appears to be correlated with the disk in their simulations. However, we also note that the DM only simulation shows a rather strikingly different radial sphericity profile indicative of a substantial influence of the baryonic material on the (central) shape: even at 10\% of the host's virial radius there is a difference of approximately 0.12 with the DM model being more aspherical. The situation actually reverses when moving even further in and below $0.05R_{\rm host}$ the dark matter component in the DM model now appears more spherical than in the SPH simulation; a clear manifestation of the baryonic mass concentration in the centre and the disk structures seen in all three host haloes, respectively. Another important feature to note is that for the host haloes we do see a difference between the shapes of the DM and total matter component as opposed to the subhaloes (cf. \Fig{fig:shapeBC}): they developed stronger and more concentrated baryon accumulations in the centre to actually affect the dark matter. This is confirmed by calculating again (though not presented) the radial dependence of the ratio of number of baryonic to total particles for the hosts: this ratio is actually approximately 40\% at $R_{\rm host}$, about 75\% at $0.1R_{\rm host}$ and even higher in the very central regions. Again, we prefer to measure the influence in ``number of particles'' as the usage of more particles entails a better sampling of (the radial dependence of) shapes. For completeness, the baryonic mass fractions at aforementioned host radii are $f_b=0.2$ at $0.1R_{\rm host}$ and $f_b=0.1$ at $R_{\rm host}$, respectively, with that fraction increasing to more than 0.8 in the central regions.

Throughout the rest of this paper we undertake the same comparison for the subhaloes, i.e. DM vs. SPH, though restricting the analysis of the SPH model to the total matter component as we do not find differences in the total and dark matter shapes (cf. \Fig{fig:shapeBC}).

\subsubsection{Radial profile of subhalo sphericity} \label{sec:shaperad}
\begin{figure}
\noindent
\centerline{\hbox{\psfig{figure=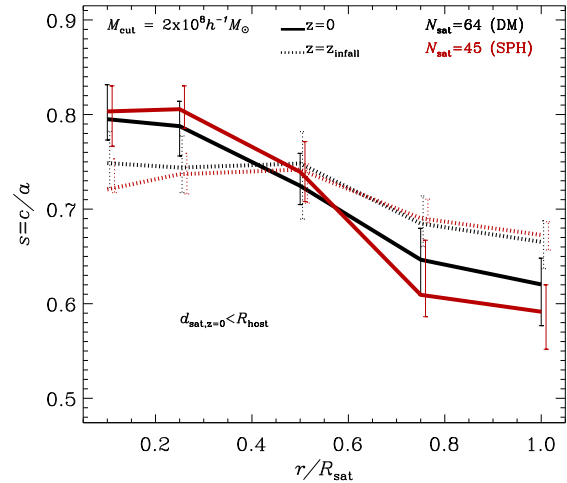,width=\hsize,angle=0}}}
\caption{The radial distribution of the medians of the sphericity $s=c/a$ based upon $a>b>c$ derived from the standard moment of inertia tensor for the DM and SPH model, respectively. The black lines correspond to the dark matter only simulation while the red lines are measures including all particles (dm, gas, and stars) in the SPH run. The solid lines refer to redshift $z=0$ while the dashed lines are based upon the subhaloes' sphericities as measured at their respective infall times.}
\label{fig:shapeS}
\end{figure}

Having established how to calculate the shapes (standard moment of inertia tensor) and which component in the SPH run to use (total matter content) we now turn to a comparison of the shapes of the DM and SPH subhaloes. In \Fig{fig:shapeS} we present the radial distribution of the median (alongside $\pm$ 25\%) values of the sphericity for the DM (black) and SPH (red) model both at redshift $z=0$ (solid lines) and at the respective infall time of the satellites (dashed lines). The most striking feature of this plot is the lack of any (statistically significant) difference between the DM and SPH runs, irrespective of the redshift. This implies that while baryon physics may affect the shapes of host haloes \citep[e.g.][and \Sec{sec:hostshapes}]{Bailin05b}, there appears to be little influence on subhaloes. Recall that we showed in \Fig{fig:shapesChosts} that we do in fact observe an influence of the baryons on the shape of the host haloes (in agreeement with Fig.~2 in \citet{Libeskind10} where we showed the influence of the baryonic component on the density profile inwards from approx. $0.1R_{\rm host}$).

However, we find that the shape evolves from the time a subhalo fell into the host and the present time: subhalos in the central parts become rounder while those in the outer parts become more apsherical . The latter is readily explained by tidal forces to be investigated in more detail in the following subsection.

\subsubsection{Dependence of sphericity on subhalo position within host}
\begin{figure}
\noindent
{\hbox{\psfig{figure=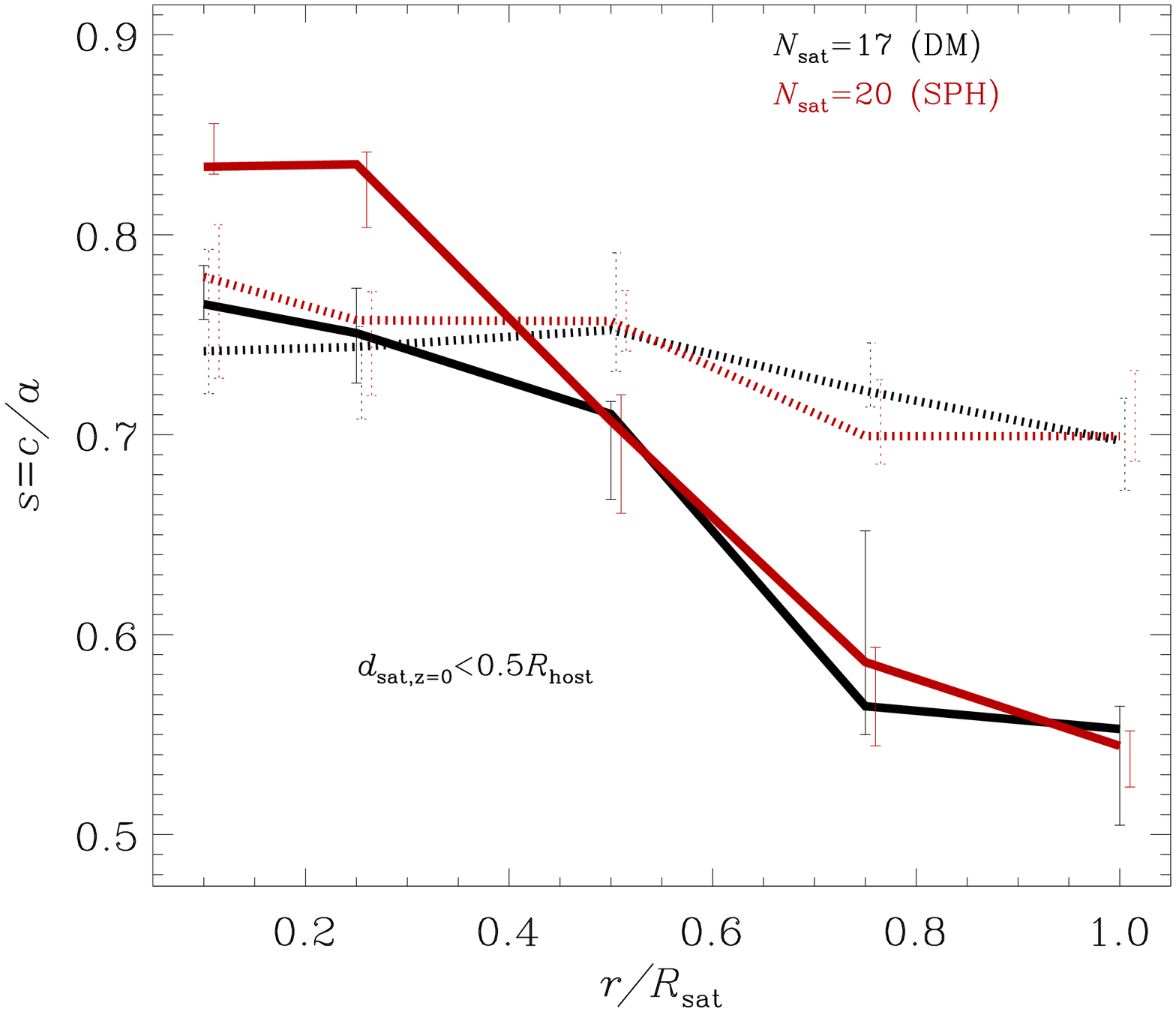,width=1\hsize,angle=0}}}
{\hbox{\psfig{figure=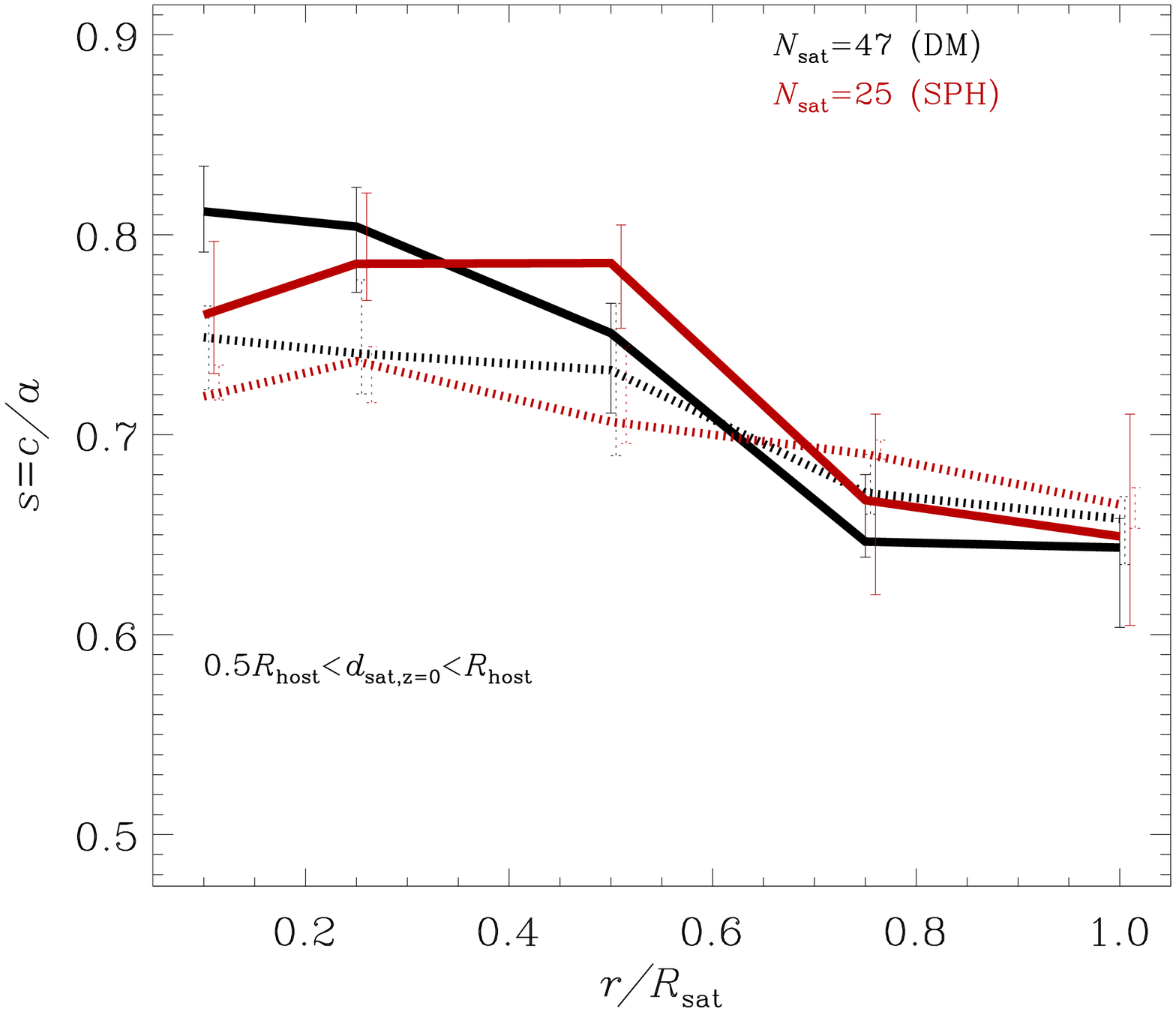,width=1\hsize,angle=0}}}
\caption{The same as \Fig{fig:shapeS} but this time dividing the subhaloes into two populations: those that end up at $z=0$ closer to the centre than 50\% of their respective host halo's radius (upper panel) and the ones between 50-100\% of the host's radius (lower panel). Note that this division is based upon the present day position $d_{\rm sat}$ of the subhalo; at infall time every satellite obviously has $d_{\rm sat}\approx R_{\rm host}$.}
\label{fig:shapeS-rad}
\end{figure}

To study the importance of tidal forces on shape, we split our sample into subhaloes closer and farther away than 50\% of the host's virial radius and present the sphericity profiles for each subset in \Fig{fig:shapeS-rad}. Note that this division is motivated by the observations of \citet{Pereira08} which indicate that the peak of the radial alignment as a function of distance from the host's centre lies at approximately half the host's virial radius (cf. their Fig.4). The upper and lower plot refer to subhaloes closer and farther away than $0.5R_{\rm host}$ at redshift $z=0$. Note that this division is based upon the distance $d_{\rm sat}$ at today's time; at infall time every satellite obviously has $d_{\rm sat}\approx R_{\rm host}$. We find that the sphericity change is primarily driven by objects closer to their respective hosts as expected if tidal forces were responsible for this result. We also observe a tendency for the closest SPH subhaloes to become more spherical than their DM counterparts in the central regions while there is a marginal opposite trend for the outer subhaloes in agreement with the findings of, for instance, \citet{Kazantzidis04b}.

\begin{figure}
\noindent
\centerline{\hbox{\psfig{figure=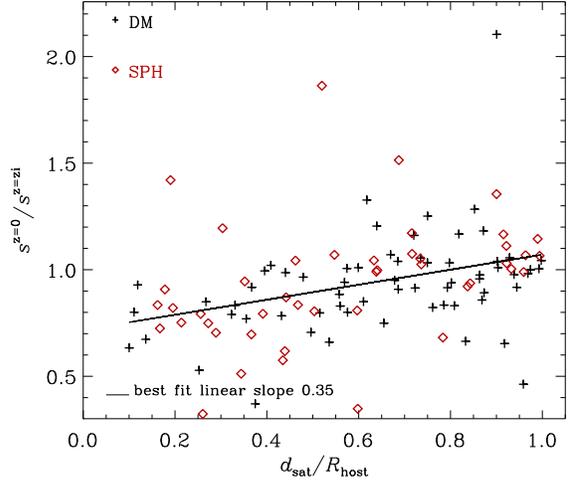,width=\hsize,angle=0}}}
\caption{The ratio of sphericities $s^{z=0}$ at redshift $z=0$ and $s^{z_i}$ at the respective infall redshift $z_i$ of the subhalo as a function of position $d_{\rm sat}$ of the subhalo at redshift $z=0$ within the host's virial radius $R_{\rm host}$. The solid line indicates the best fit linear relation for the combined sample (i.e. DM and SPH together).}
\label{fig:shapeSdist}
\end{figure}

To better quantify the sphericity change (as measured for all our subhaoes at their respective radii $R_{\rm sat}$) we plot in \Fig{fig:shapeSdist} the ratio $s^{z=0}/s^{z=z_i}$ against the position $d_{\rm sat}$ of the subhalo within the host, where $s^{z=0}$ is the sphericity as measured today and $s^{z=z_i}$ at infall time. This ratio is primarily below unity indicating a drop in sphericity. Further, subhaloes ending up closer to their host's centre show a (marginally) stronger drop in agreement with the picture of tidal origin. This correlation, (i.e. stronger sphericity evolution for subhaloes closer to their host) is confirmed by the Spearman rank coefficient taking a value of 0.46 for both the combined DM and SPH model and a linear fit to the combined data giving a slope of $0.35$ (indicated by the solid line).

\subsection{The Radial Alignment}
\label{sec:radialalignment}
As we have just seen, subhaloes/satellites themselves are -- just like field haloes -- aspherical systems as previously shown by \citet[e.g.][]{Knebe08}. Furthermore, it has also been reported in the literature that the orientation of their shapes is not random and their major axis preferentially points towards their host centre \citep[e.g.][]{Kuhlen07, Faltenbacher08, Pereira08, Knebe08, Knebe08b}. This ``radial alignment'' is quantified by the (cosine of the) angle between the position vector of the subhalo ${\bf d}_{\rm sat}$ in the rest frame of its host halo and the major axis ${\bf a}_{\rm sat}$ of the actual subhalo (again, as determined via the eigenvectors of the standard moment of inertia tensor). This quantitiy

\begin{equation} \label{eq:radialalignment}
\cos {\phi} = {\bf d}_{\rm sat} \cdot {\bf a}_{\rm sat} 
\end{equation}

\noindent
will be the target of our study in the following subsection. in order to be able to properly measure the radial alignment, and in compliance with previous studies \citep[cf.][]{Pereira08,Knebe08}, we wish to further restrict ourselves to those subhaloes with well defined major axes, pruning our sample such that all subhaloes have an intermediate to long axis ratio of $b/a<0.9$. Please refer to \Tab{table2} for the number of subhaloes in this restricted sample.

\subsubsection{Radial profile of radial alignment}
\begin{figure}
\noindent
\centerline{\hbox{\psfig{figure=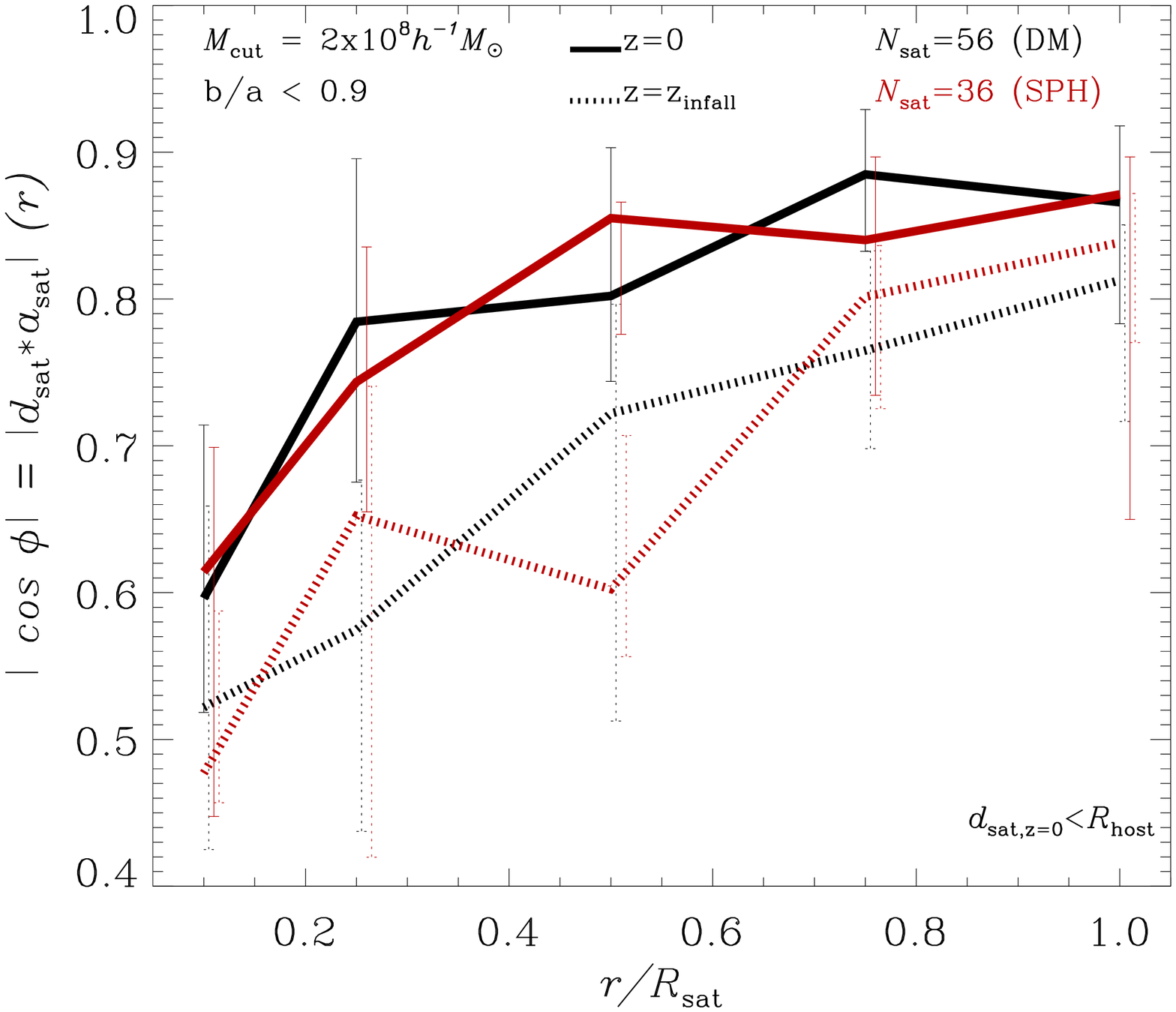,width=\hsize,angle=0}}}
\caption{The radial distribution of the medians of the radial alignment based upon $a>b>c$ derived from the standard moment of inertia tensor. The black lines correspond to the dark matter only simulation while the red lines are measures including all particles (dm, gas, and stars) in the SPH run. The solid lines refer to redshift $z=0$ while the dashed lines are based upon the subhaloe's sphericities at their respective infall times.}
\label{fig:radali}
\end{figure}

The radial alignment (and its evolution since infall time) is shown in \Fig{fig:radali} where the radial distribution of the medians of $|\cos {\phi}|$ as defined by \Eq{eq:radialalignment} is plotted. We note that while the signal appears to be strong (as expected) there is no noticeable difference between the DM and SPH model. 
This comes as no surprise as we previously mentioned that there is virtually no difference between the shapes of subhaloes in the DM and the SPH model (cf. \Sec{sec:shapes}). That said, we note the trend for the signal to strengthen over time -- akin to the observation of the evolution of sphericities. This is expected if the signal stems from evolutionary rather than enviromental effects as shown by \citet{Kuhlen07, Pereira08, Knebe08}. We further find that the signal weakens when moving the point of reference for the subhalo shape inside, just as already indicated by \citet{Kuhlen07}. However, this cannot be related to the fact that the subhaloes become more spherical in the inside (cf. \Fig{fig:shapeS}) as (though not explicitly shown here) the radial aligment follows a similar radial trend throughout the subhalo when using the reduced moment of inertia tensor.

\subsubsection{Dependence of radial alignment on subhalo position within host}
\begin{figure}
\noindent
{\hbox{\psfig{figure=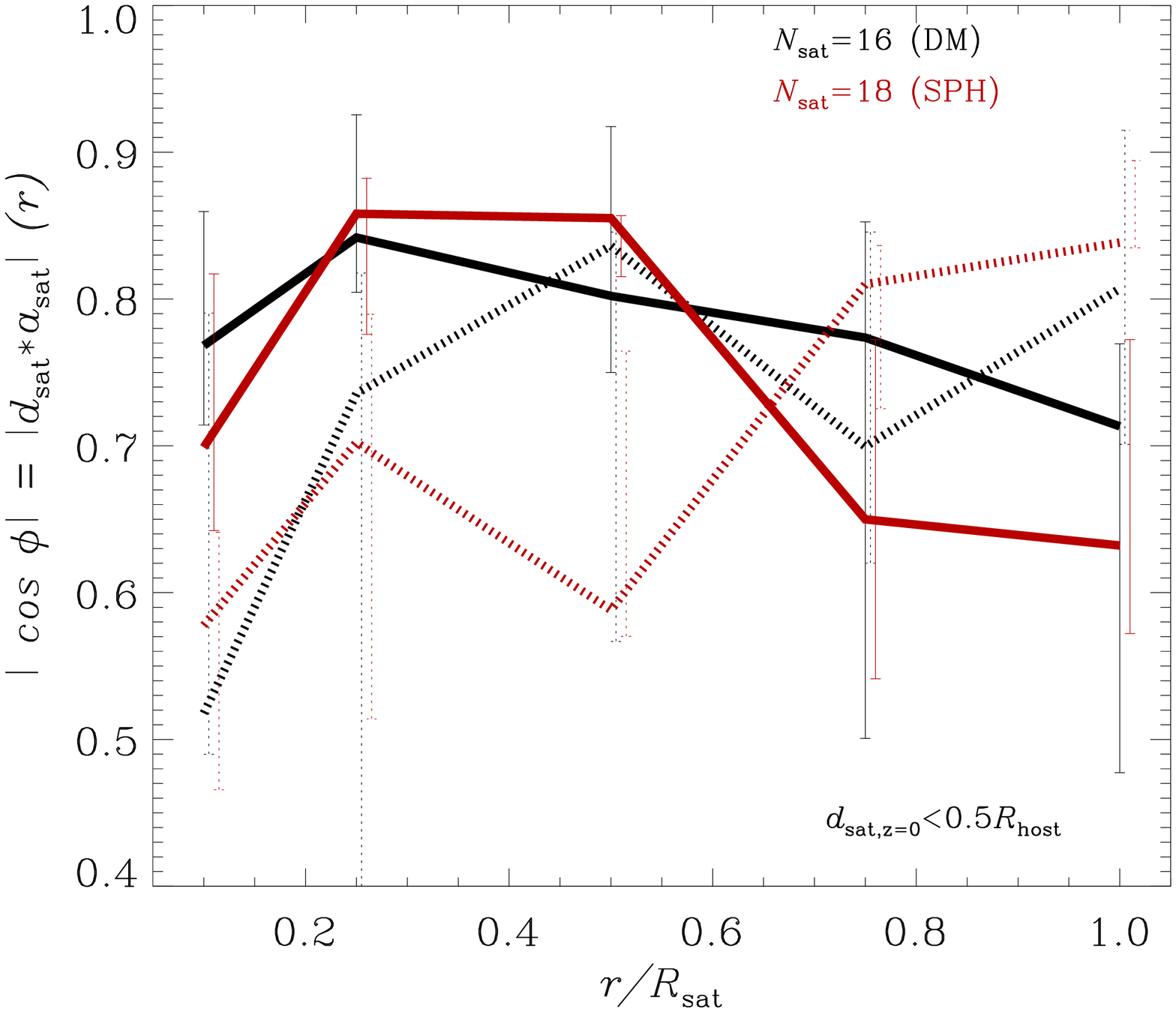,width=\hsize,angle=0}}}
{\hbox{\psfig{figure=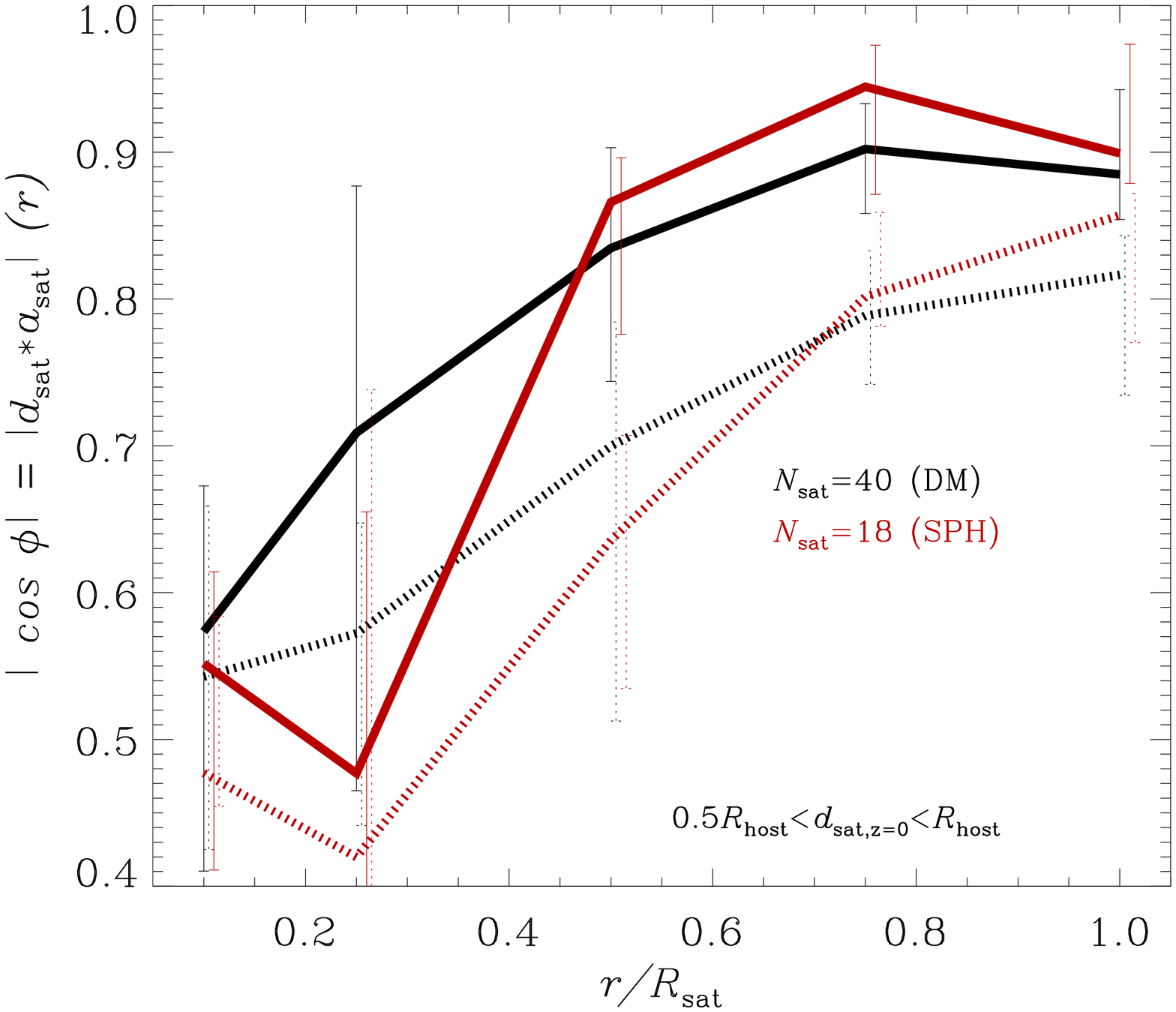,width=\hsize,angle=0}}}
\caption{The same as \Fig{fig:radali} but this time dividing the subhaloes into two populations: those that end up at $z=0$ closer to the centre than 50\% of their respective host halo's radius (upper panel) and the ones inbetween 50-100\% of the host's radius (lower panel).}
\label{fig:radali-rad}
\end{figure}

As before, we divide our subhaloes into two distinct samples according to position in the halo: objects closer to the centre than 50\% of their host's virial radius and satellites between 50 and 100\% of the host's radius. This division is shown in \Fig{fig:radali-rad} where the distribution of the radial alignment for the inner (outer) sample can be viewed in the upper (lower) panel. We find that the signal primarily evolves for those subhaloes that remain in the outskirts of the halo: subhaloes that end up closest to the host's centre at redshift $z=0$ experience a mild weakening of the radial alignment  - in fact the evolution is also consistent with no evolution due to the large scatter. At first this may appear surprising as we previously showed that the shapes become more aspherical over time primarily for those objects ending up closer to the host. One may thus expect a strengthening of the radial alignment. But as we show later, the scenario is a bit more complicated and is related to the fact that subhaloes move faster at pericentre. However, our findings are consistent with \citet{Kuhlen07} and \citet{Pereira08} who found a considerably stronger signal for satellites between $[0.5,1.0]R_{\rm host}$ than for objects closer than this. These authors explain this by asserting that the subhalo is moving too fast at pericentre for tidal torquing to be effective \citep[cf. Fig.8 in][]{Pereira08}. Below, in subsection~\ref{sec:velocity}, we provide direct evidence for this scenario.

\begin{figure}
\noindent
{\hbox{\psfig{figure=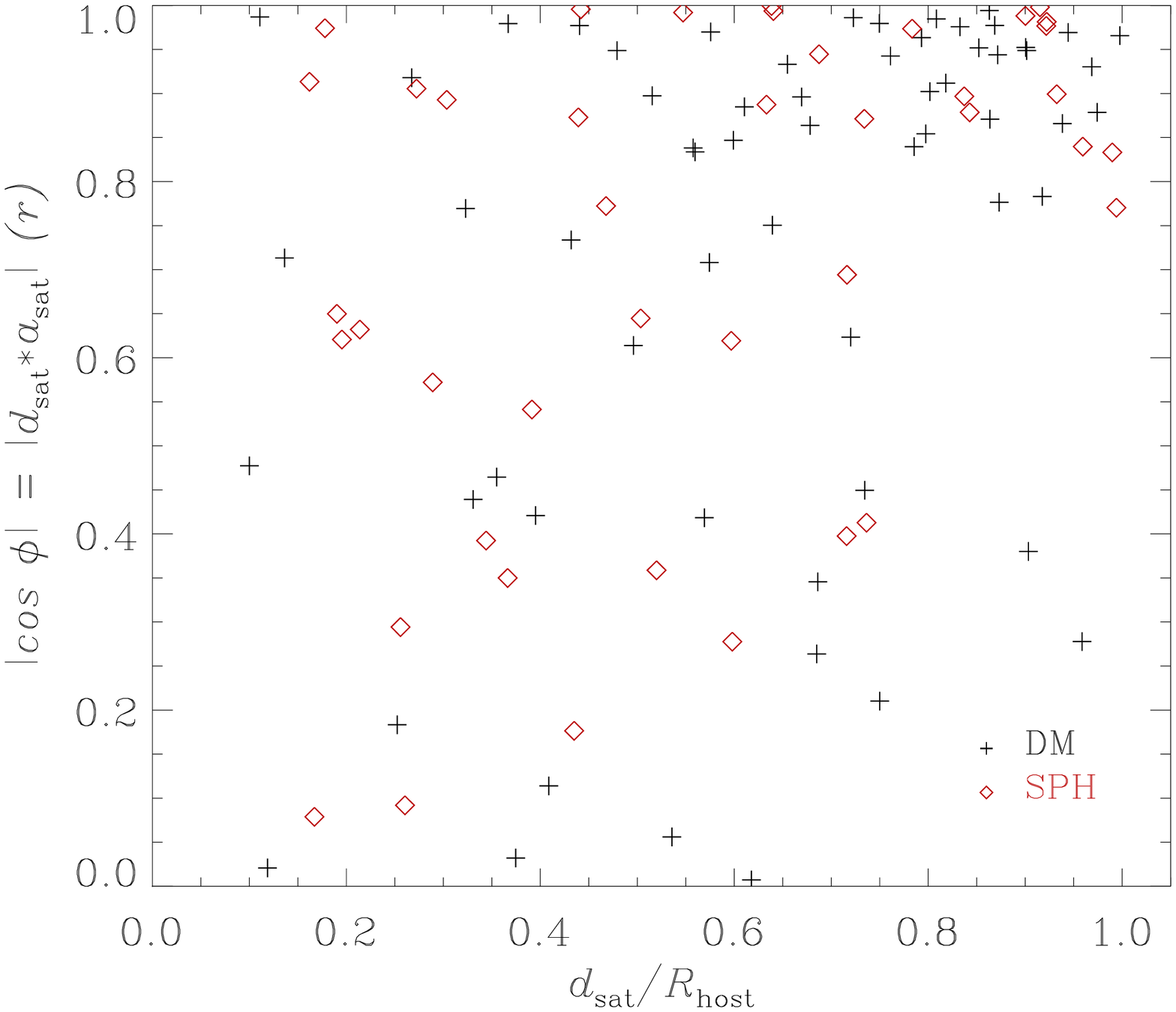,width=1\hsize,angle=0}}}
{\hbox{\psfig{figure=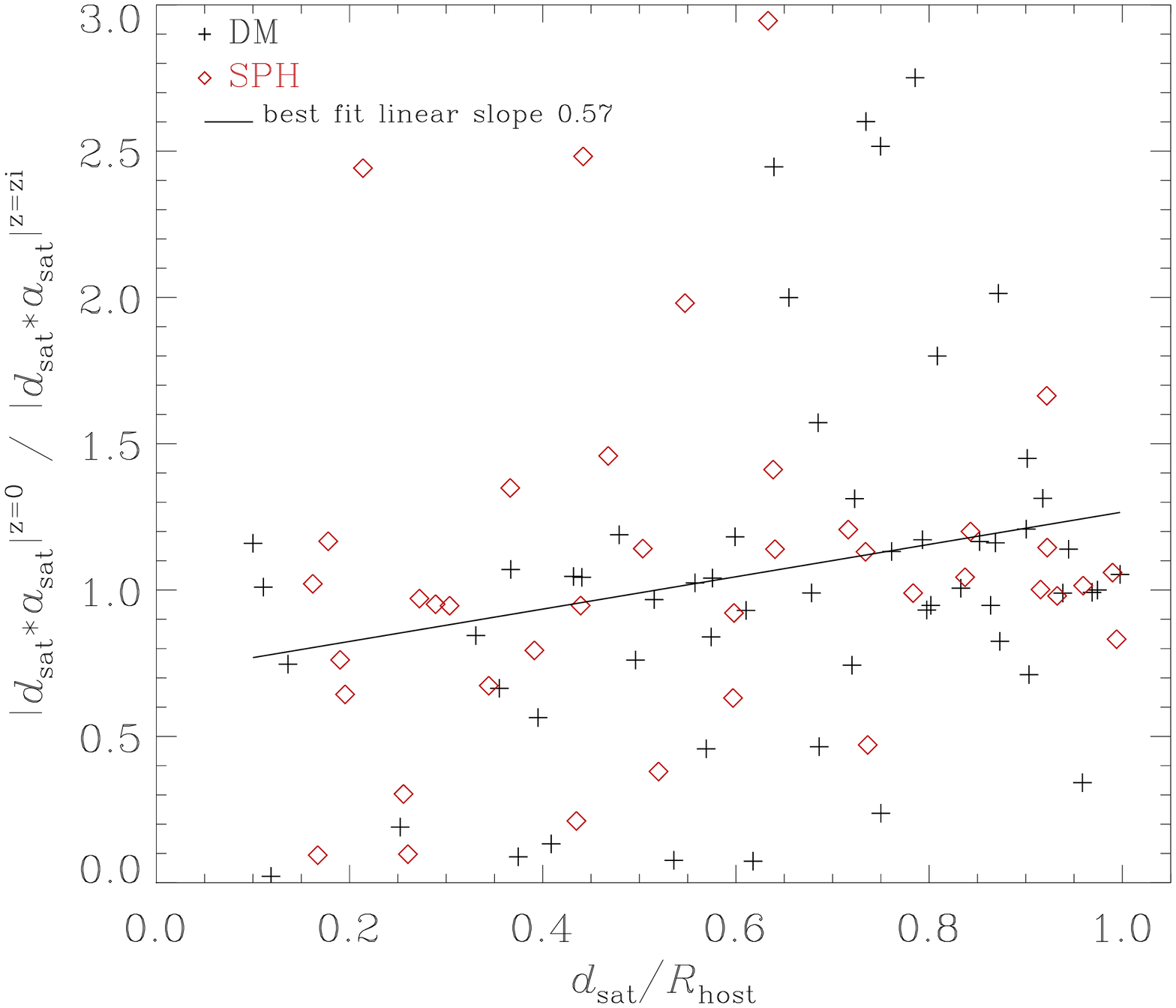,width=1\hsize,angle=0}}}
\caption{The radial alignment as given by \Eq{eq:radialalignment} as a function of the subhalo's position $d_{\rm sat}$ within its host halo. The upper panel shows the relation for redshift $z=0$ while the lower panel shows the ratio between the value at redshift $z=0$ and the respective infall redshift of the subhalo.}
\label{fig:radalidist}
\end{figure}

We conclude this subsection with an investigation of the correlation between the radial alignment and subhalo position on an individual object basis. To this extent, we present in \Fig{fig:radalidist} the radial alignment $\cos {\phi} = {\bf d}_{\rm sat} \cdot {\bf a}_{\rm sat}$ for each subhalo vs. its position $d_{\rm sat}$ as measured today (upper panel) and the ratio between ${\bf d}_{\rm sat} \cdot {\bf a}_{\rm sat}$ as measured today and at infall time (lower panel). We observe that the radial alignment at $z=0$ shows a (marginal) tendency to be stronger for objects in the outskirts of the host halo -- as confirmed by reading off the corresponding numbers for the medians at $r/R_{\rm sat}=1$ in the upper and lower panel of \Fig{fig:radali-rad}. The lower panel of \Fig{fig:radalidist} further confirms a strengthening of the signal over time -- at least for the subhaloes residing in the outer parts of their host halo at redshift $z=0$ as noted earlier. This correlation between evolution of radial alignment and subhalo position in the host is again confirmed by the rather large value of 0.74 for the Spearman rank coefficient as well as the best fit linear slope of 0.57. Please note that the lower panel also indicates a (mild) weakening of the radial alignment for objects ending up close to the host in agreement with the picture that the signal did not have enough time to adjust itself for such ``speeding'' subhaloes. We shall have a closer look at exactly this phenomenon in the subsequent Subsection~\ref{sec:velocity}.

\subsection{Correlations with Subhalo Velocity}
\label{sec:velocity}
One explanation for the stronger radial alignment in the outer parts is tidal torquing and the fact that subhaloes are simply moving too fast at  pericentre for the radial alignment to become effective \citep[e.g.][]{Pereira08}. To confirm this picture we examine the correlation of the long axis ${\bf a}_{\rm sat}$ with the subhalo's velocity ${\bf v}_{\rm sat}$ in the host halo's rest frame:

\begin{equation}
 \cos {\phi} = {\bf v}_{\rm sat} \cdot {\bf a}_{\rm sat} \ .
\end{equation}

\begin{figure}
\noindent
{\hbox{\psfig{figure=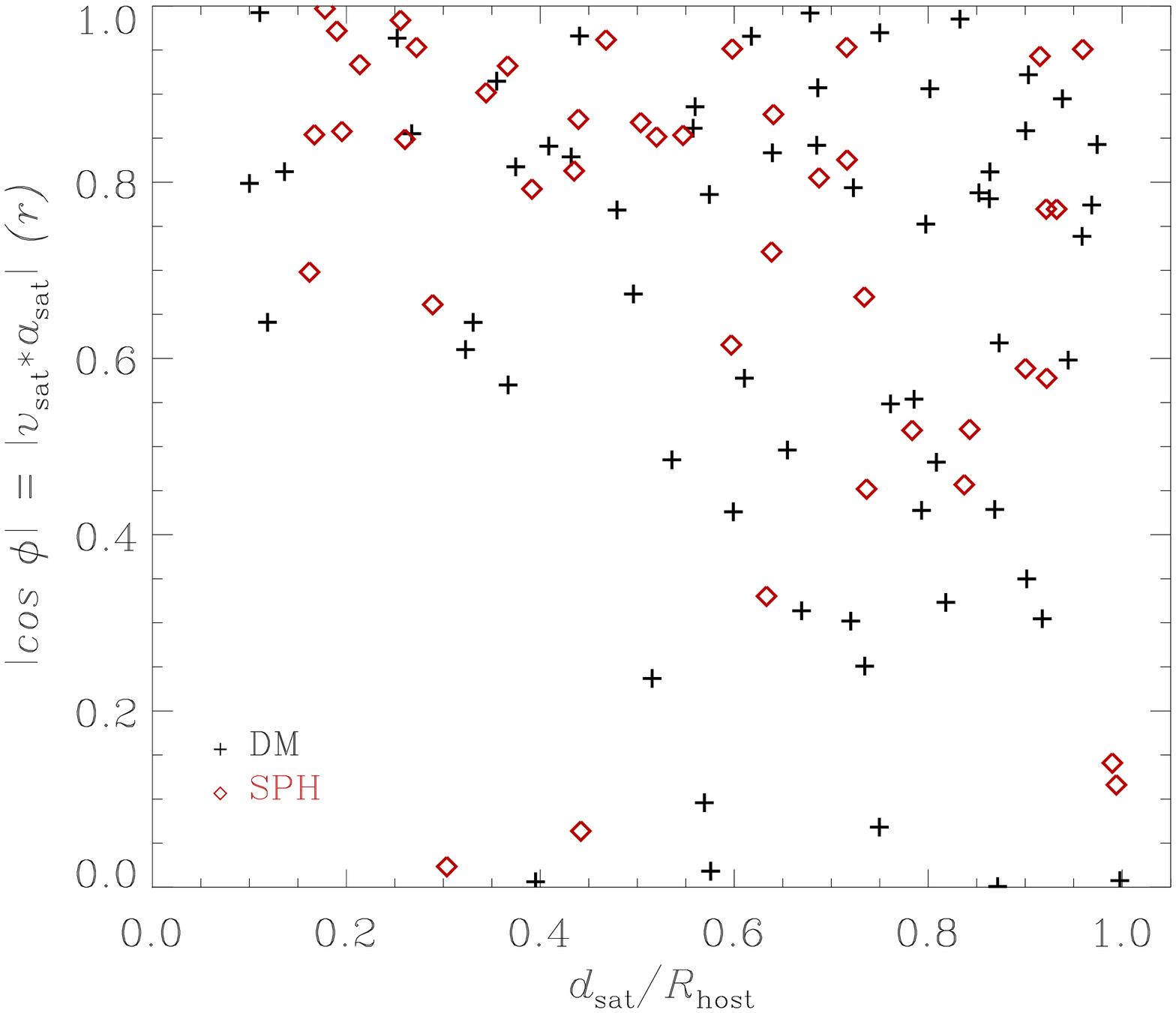,width=\hsize,angle=0}}}
{\hbox{\psfig{figure=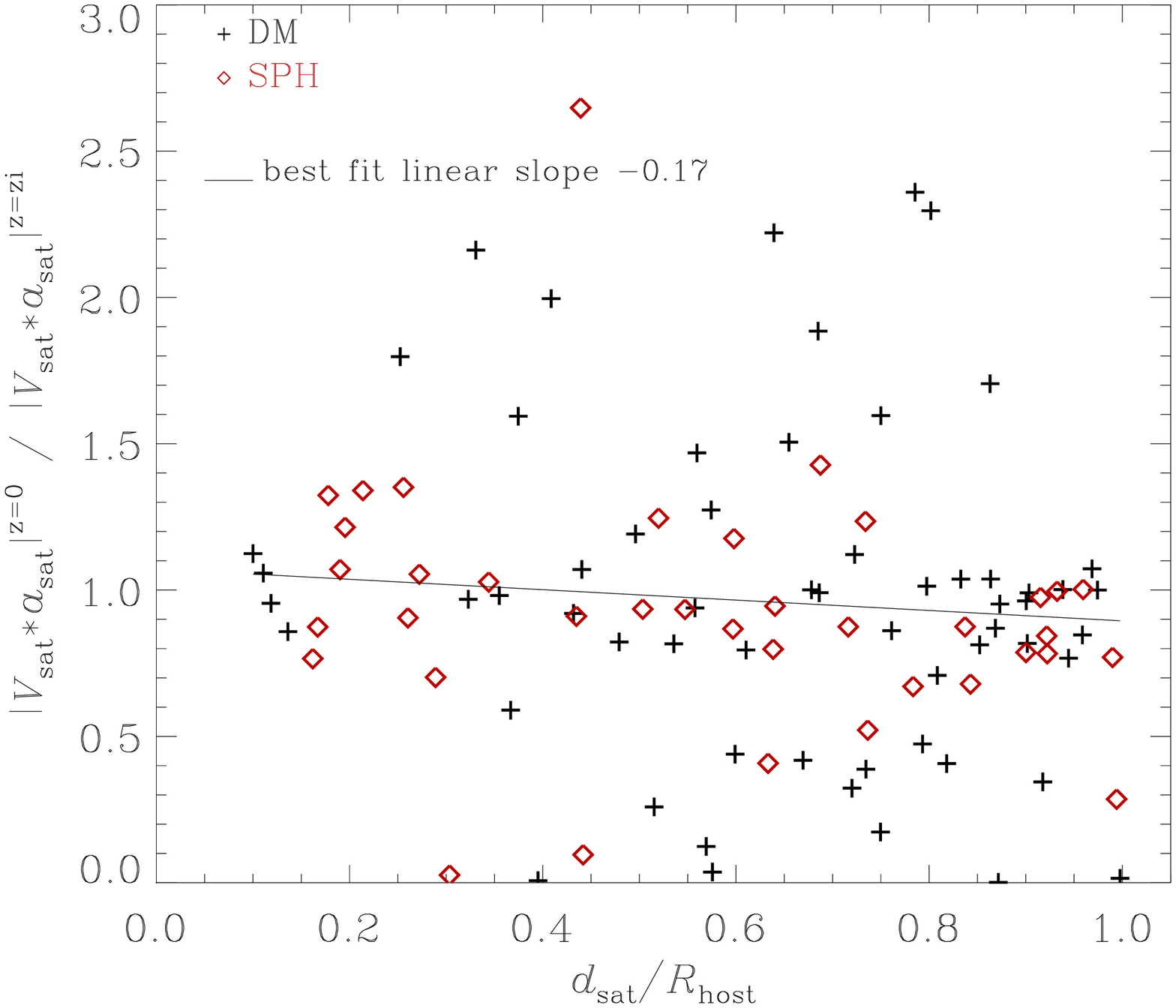,width=\hsize,angle=0}}}
\caption{The correlation between the major axis ${\bf a}_{\rm sat}$ of the subhalo (as measured at its radius $R_{\rm sat}$) and the velocity ${\bf v}_{\rm sat}$ in the rest frame of the respective host halo as a function of position $d_{\rm sat}$ of the subhalo within its host. The upper panel shows the relation for redshift $z=0$ while the lower panel shows the ratio between the value at redshift $z=0$ and the respective infall redshift of the subhalo.}
\label{fig:EsatVsat}
\end{figure}

\Fig{fig:EsatVsat} shows the modulus of $\cos {\phi}$ as a function of the distance $d_{\rm sat}$ to the host centre both at redshift $z=0$ (upper panel) and the ratio between ${\bf v}_{\rm sat} \cdot {\bf a}_{\rm sat}$ as measured today and at infall time (lower panel). We note that subhaloes ending up in the central regions of their hosts have their shapes (as measured by the major axis ${\bf a}_{\rm sat}$ at $R_{\rm sat}$) aligned with the velocity vector ${\bf v}_{\rm sat}$. This explains why we saw a weakening of the radial alignment signal in \Sec{sec:radialalignment}: \textit{the velocity vector is tangential to the orbit and hence not pointing towards the host}. However, the shape appears to be elongated along the flight path (which we have confirmed by looking at ${\bf d}_{\rm sat} \cdot {\bf v}_{\rm sat}$ -- though not explicitly presented here -- where we found a (mild) correlation at infall time) and hence the correlation of both quantities. This observation is naturally in agreement with the previous explanation of \citet{Pereira08}: the subhalo does not have enough time to re-arrange its shape at pericentre passage; it nevertheless gets stretched/tidally torqued when passing close to the host centre. However, it appears that this signal is already present at infall time, i.e. even the infall shape and infall velocity are more strongly correlated for those subhaloes ending up closer to the host which is confirmed by the rather weak evolution of ${\bf v}_{\rm sat} \cdot {\bf a}_{\rm sat}$ as seen in the lower panel of \Fig{fig:EsatVsat}. There we further find only a mild dependence of this evolution with position $d_{\rm sat}$ within the host as confirmed by both the best fit linear slope of $-0.17$ and the Spearman rank coefficient of $-0.28$ for the combined sample.

We like to close with a cautionary remark: from both \Fig{fig:radalidist} and \Fig{fig:EsatVsat} one may be inclined to infer that the (infall) velocity of subhaloes should be randomly distributed as there appears to be no correlation between shape (as measured by ${\bf a}_{\rm sat}$) and ${\bf v}_{\rm sat}$ whereas the shape is correlated with the position vector ${\rm d}_{\rm sat}$ for the subhaloes ending up in the outer parts of their respective host halo. However, one needs to bear in mind that subhaloes on radial orbits (i.e. ${\bf v}_{\rm sat}$ aligned with ${d}_{\rm sat}$) preferentially end up in the central regions; subhaloes lingering in the outer parts are favourably on more circular orbits. This picture is confirmed by actually plotting ${\bf v}_{\rm sat} \cdot {\bf d}_{\rm sat}$ which we omitted here: those subhaloes who had both these vectors aligned at infall time (i.e. radial orbits) actually end up closer to the centre with smaller $d_{\rm sat}/R_{\rm host}$; the ones remaining in the outskirts show a weaker alignment at infall time.

\section{Summary and Conclusions}
\label{sec:summary}
In this paper, we use two simulations performed within the CLUES project to study both the shape and radial alignment of subhaloes; one of the simulations is a dark matter only model while the other run includes all the relevant gas physics and star formation recipes.

We first set out to determine the most accurate method to measure the shape of subhaloes. To this end we compared two similar methods: the standard and reduced moment of inertia tensor. We found that due to the $1/r^2$-weighing, the reduced moment of inertia tensor gives a stronger bias towards the central shape of subhaloes. Since in this work we are interested in the effects of tidal forces that primarily act in the outer parts, we decided to employ the standard moment of inertia tensor. We further investigated the dependence of measuring the sphericity of SPH subhaloes on the individual components, e.g. solely the dark matter particles or all particles including gas and stars. Despite the presence of baryonic particles in our SPH subhaloes,  we are unable to find any significant differences in the sphericities of subhaloes when calculated using either all the particles or just the dark matter  ones. We thus decided to use all particles when determining the shape of SPH subhalos. This does not automatically imply that baryons have no effect on shape just that in an SPH subhalo the baryonic and dark matter components follow the same radial shape profile.

Our first result is that -- when comparing the sphericities of subhaloes in the DM and the SPH model -- the distribution of shapes is not affected by including baryons. We further note that we also investigated the triaxiality in the same way as the sphericity  and were led to the same conclusions; therefore, we use the terms sphericitity and shape interchangeably. However, there is (for both models) an evolution in time: subhaloes become more aspherical over time with this evolution being driven primarily by objects closer to the centre of the host, as expected if tidal forces were responsible for alterations to the shape.

We further find that the radial alignment is not affected by gas physics, which should come as no surprise since we did not find an influence of gas physics on the shapes of subhalos. We thus confirm the picture of an evolutionary (rather than environmental) origin of the signal: there is considerable evolution of the radial alignment since infall of a subhalo with the signal being weaker for objects close to the centre. This weakening can be explained by the fact that the subhaloes move too fast at pericentre for the alignment signal to adjust. This is confirmed by a correlation between the shape of the subhalo and its velocity vector, the tangent to the actual orbit not pointing towards the host centre.

The most remarkable outcome of this study though remains that the inclusion of gas physics has no affect on the (dark matter) shapes (and hence radial alignment) of subhaloes in cosmological simulations of structure formation. While we do find an influence in our suite of host haloes along the lines reported by other authors \citep[e.g.][]{Dubinski94, Bailin05b, Gustafsson06, Debattista08, Abadi09, Tissera09}, there appears to be no impact on subhaloes. Why are subhaloes different to host haloes in that respect? After all, the inclusion of baryons should affect the dark matter, too: \cite{Blumenthal86} showed that dissipative baryons will lead directly to the adiabatic contraction of haloes \citep[see also][]{Gnedin04} and should thus be a critical ingredient in determining halo and subhalo properties. However, here we find that while it most certainly affects the host haloes (cf. \Fig{fig:shapesChosts}) it appears unimportant for the subhaloes. We though need to note that this paper looks primarily at low-mass subhaloes (Milky Way dwarf galaxies), and hence one could imagine the result changing significantly on cluster scales, where the substructure now corresponds to massive galaxies, where the distribution and affect of baryons might be quite different.

Although we are confident that our results are not driven by numerical artifacts (see \Sec{sec:howto}), we note that the fraction of baryonic particles in the host haloes is substantially larger than in our subhaloes: 20-30\% of the particles in a given subhalo are baryonic particles. This ratio increases with decreasing subhalo radius: the baryon number fraction is approximately 40\% at $R_{\rm host}$, $\sim$75\% at $0.1R_{\rm host}$ and even higher in the very central regions.  Note that we chose to adhere to ``number fractions'' as a a greater number of particles will allow for a better sampling of the (radial dependence of) (sub-)halo shape -- irrespective of their mass; the corresponding baryonic mass fractions can be found in \Sec{sec:howto} and \Sec{sec:hostshapes}, respectively. Given this difference in host and subhalo fractions we like to caution that even higher resolution may be required to confirm the results presented here. We further checked whether the baryons in the subhaloes are primarily in the form of gas or stars. In that regards we found that the gas mass fraction is substantially lower than the stellar mass fraction. In fact, the median gas to stellar mass fraction is of order 5\%, i.e. 95\% of the baryons are stars. Under the assumption that those stars have formed prior to the infall of the subhalo into its host and the fact that star particles are effectively behaving like dark matter particles we may draw the conclusion that our results are not as suprising as one may think: even though the stars have a different formation history, i.e. they are born out of gas which had the option to cool and sink to the centre of the subhaloes' potential wells, the influence of the baryonic/stellar content of subhaloes will be limited as observed in this paper. However, the study presented here focused on the shape of the dark matter only component. To further investigate the influence of the baryonic component we will in an upcoming paper, study the shapes of the baryonic component (Libeskind et al., in preparation).

\section*{Acknowledgements}
AK is supported by the Ministerio de Ciencia e Innovacion (MICINN) in Spain through the Ramon y Cajal programme. NIL is supported by the Minerva Stiftung of the Max Planck Gesellschaft. SRK acknowledges support by the MICINN too under the Consolider-Ingenio, SyeC project CSD- 2007 -00050. We thank DEISA for granting us supercomputing time on MareNostrum at BSC and in SGI- Altix 4700 at LRZ, to run these simulations under the DECI- SIMU-LU and SIMUGAL-LU projects.  We also thank ASTROSIM for giving us different travel grants to visit our respective institutions.  GY acknowledges financial support from MEC (Spain) under projects FPA2006-01105 and AYA2006-15492-C03. YH has been partially supported by the ISF (13/08).

\bibliography{archive} \bsp

\label{lastpage}

\end{document}